\documentclass[twocolumn]{autart}    % Enable this line and disable the 
\usepackage{graphicx}          
\usepackage[cmex10]{amsmath}[12pt]
\usepackage{url}
\usepackage{amssymb}
\usepackage{algpseudocode}
\usepackage{color}
\usepackage{algorithm}
\usepackage{cite}
\usepackage{picins}

\newcommand{\R}{\ensuremath{\mathbb{R}}}
\newcommand{\N}{\ensuremath{\mathbb{N}}}
\newcommand{\spann}{\ensuremath{\operatorname{span}}}

\newcommand{\dist}{\ensuremath{\operatorname{dist}}}

\newcounter{enumctr}
\newenvironment{enum}{\begin{list}{(\roman{enumctr})}{\usecounter{enumctr}}}{\end{list}}

\begin{document}

\begin{frontmatter}

\title{On the stability and convergence of a class of consensus systems with a nonlinear input \thanksref{footnoteinfo}} % Title, preferably not more 

\thanks[footnoteinfo]{This paper was not presented at any IFAC 
meeting. Corresponding author M. Liu. Tel.: +353 871760843}

\author[mm]{Mingming Liu}\ead{mingming.liu@ucd.ie},    % Add the 
\author[fb]{Fabian Wirth}\ead{fabian.lastname@uni-passau.de},               % e-mail address 
\author[mt]{Martin Corless}\ead{corless@purdue.edu},  % (ead) as shown
\author[mm]{Robert Shorten}\ead{robert.shorten@ucd.ie}

\address[mm]{School of Electrical, Electronic, and Communications Engineering, University College Dublin, Dublin, Ireland}
\address[fb]{the Faculty of Computer Science and Mathematics, University of Passau, Passau, Germany}             % full addresses
\address[mt]{ School of Aeronautics and Astronautics, Purdue University, West Lafayette, IN 47907-2045, USA}        % here.

                             % keyword list or with the 
                                          % help of the Automatica 
                                          % keyword wizard

\begin{abstract}                          % Abstract of not more than 200 words.
We consider a class of consensus systems driven by a nonlinear input.  Such systems arise in a class of IoT applications. Our objective in this paper is to determine conditions under which a certain 
partially distributed system converges to a Lur'e-like scalar system, and to provide a rigorous proof of its stability. Conditions are derived for the non-uniform convergence and stability of such a system and an example is 
given of a speed advisory system where such a system arises in real engineering practice. 
\end{abstract}

\end{frontmatter}

\section{Introduction}
We consider   nonlinear discrete-time systems described by
\begin{eqnarray}
\label{eq:base_eqn}
x(k+1) = P(k)x(k) + \mu \big(r - g(x(k))\big)e
\end{eqnarray}
where $k=0,1,2, \dots,$ $x(k) \in \mathbb{R}^{n}$, $P(k)$ is a $n \times n$ row stochastic matrix, $e = [1\ 1\, \dots\,1]^T \,$, $\mu$ and $r$
are scalars while $g$ is a scalar valued function. Equation (\ref{eq:base_eqn}) arises in  consensus problems subject to an output constraint. It basically says that if consensus is achieved, it must be achieved subject to the equilibrium constraint $g(x^*) = r$. That is, at equilibrium
\begin{equation}
g(x^*) = r,  
\end{equation} 
with $x_i^* = x_j^*$ for all  $ i,j \in \left\lbrace 1,2,...,n \right\rbrace$. Equation (\ref{eq:base_eqn}) is of interest as it arises in many situations in the study of the {\em internet  of  things} (IOT). For example, in some situations a group of agents are asked to achieve a fair allocation of a constrained resource; TCP is an algorithm that strives to achieve this objective in internet congestion control. Recently, similar ideas have been applied in the context of the charging of electric vehicles, smart grid applications, and in the regulation of pollution in an urban context \cite{twinlin,IEVC2014}. A second application arises when one wishes to optimise an objective function subject to certain privacy constraints. For example,  collaborative cruise control systems are emerging in which a group of vehicles on a stretch of road share information to determine a common speed  that minimises fuel consumption of the group subject to some constraint (traffic flow, pollution constraints) \cite{ITS}. 
Since each car is individually optimised for a potentially different speed, the technical challenge is for the group of cars to agree on a common speed without an individual revealing any of its inner workings to other vehicles. Other examples of this nature abound. As we have mentioned, for reasons of privacy, usually one does not attempt to solve such problems in a fully distributed manner. Neither, for reasons of robustness, scale, and communication overhead, does one attempt to solve them in a centralised manner. Rather, one uses a mix of local communication, and limited broadcast information, to solve these problem in a manner that conceals the private information of each of the individual agents. Implicit and explicit consensus algorithms that exploit local and global communication strategies are proposed and studied in \cite{martin2,knorn2011results}. Equation (\ref{eq:base_eqn}) is perhaps the simplest algorithm of the explicit consensus algorithm with inputs, admitting a very simple intuitive understanding.
It is well known that a row stochastic matrix $P$ operates on a vector $x \in \R^{n}$ such that 
$\textrm{max}(x) - \textrm{min}(x) \geq \textrm{max}(Px) - \textrm{min}(Px)$
where $\textrm{max}(x)$ and $\textrm{min}(x)$ are defined as the maximum and minimum component in vector $x$, respectively. Since the addition of $\big(r - g(x(k))\big)e$ does not affect this contraction, intuition suggests that $x_i(k) -x_j(k)\rightarrow 0$
as $k$ increases and eventually, the dynamics of (\ref{eq:base_eqn}) will be governed by the following scalar {\em \bf Lur'e system}: 
\begin{eqnarray}
\label{eq:lure}
y(k+1) = y(k) + \mu(r\!-\!g(y(k)e)), 
\end{eqnarray}
with $x_i(k) = y(k)$ asymptotically for all $i$. Intuition  further suggests that, as long as (\ref{eq:lure}) is stable, then so is (\ref{eq:base_eqn}).  
Arguments along these lines, in support of (\ref{eq:base_eqn}), are given in \cite{martin2,knorn2011results}. However, these arguments are not complete
in the sense that certain important properties are assumed to hold true.
Our objective therefore in this brief note is to establish
conditions on the function $g$  for which global uniform asymptotic
stability is assured, and to rigorously prove the resulting assertions.

\subsection{Comments on related literature} 
Before proceeding it is prudent to discuss connections to related work. 
\begin{itemize}
	\item{(i) Cascade Systems :} The setup we study can be viewed as a more general
	case of the systems studied in \cite{lu2007synchronization} 
	where the authors prove local synchronization results for a general
	class of nonlinear time-varying systems. 
	% In contrast to the assumptions
	% of that paper we require fewer differentiability assumptions and state
	% conditions which ensure global convergence.
	However, in contrast to the assumptions of that paper we do not require
	differentiability of the system maps and we state conditions which ensure
	global convergence. Further related work concerns stability analysis of
	nonlinear cascades \cite{loria2002stability,loria2003uniform,nevsic2004uniform}; we will
	comment on the relation to this literature in the system description and
	when stating results.  \newline
	\item{(ii) Consensus :} The setup we discuss is obviously connected 
	to work on consensus. While the literature on consensus is too rich to give a complete survey
	here, recent surveys are available in 
	\cite{Liter2,mesbahi2010graph}. Here we briefly note that 
	the standard problem studied in this
	literature are conditions that guarantee convergence of solutions to
	the consensus subspace. In this paper, we study the more specific
	problem of convergence to a specific point in which consensus is
	reached. Standard results ensuring consensus will therefore not apply
	in any classical sense to the problem studied here. Furthermore, a 
	standard assumption in the case of a consensus system is a
	connectivity assumption on the communication graph. For a discussion
	of conditions used in this area we refer the reader to
	\cite{Liter3,blondel2005convergence,mesbahi2010graph}.\newline
	\item{(iii) Optimisation :} As will be seen later, one particular
	application for the class of systems discussed
	in this paper allows for the solution of distributed
	optimisation problems using a consensus approach.
	While this is just one application of our result, and
	a minor part of this paper, some comments placing
	our work in this context are in order. Note, the idea
	to use consensus techniques to obtain approximations
	of optimal solutions has already been studied in other papers; see 
	\cite{nedic2010constrained,shi2013reaching,burger2014polyhedral} for some work in this direction. For example, in \cite{burger2014polyhedral}, a constraint exchange
	based consensus algorithm was proposed, where
	the authors combined the ideas with dual decomposition
	and cutting-plane methods to solve convex and
	robust distributed optimisation problems via polyhedral
	approximations. While this approach is applicable to 
	more general problems than those studied here, 
	it is also more complex. Further, we note that
	many of the other techniques in the literature
	rely on the use of individual constraint sets for
	the agents and projections onto that set. These reduce
	to standard consensus in the case that no constraints
	are present. \newline
	\item{(iv)  Convergence :} As a further difference we point out that we give convergence results which can be non-uniform, whereas the authors in \cite{nedic2010constrained} point out that they rely on uniform convergence. 
\end{itemize}

\section{Notation, Conventions and Preliminary Results}

\subsection{Notation} \label{sec:notation}

We denote the standard basis  in $\R^n$ by  the vectors $e_1, \ldots, e_n$.
% and the vector with all entries equal to one by
Note that $e = \sum_{i=1}^n e_i$. 
A matrix $P \in \R^{n \times n}$ is called {\sf row stochastic}, if all its entries are nonnegative and if all its row sums  equal  one.
The row sum condition is equivalent to $Pe = e$, that is,
$e$ is an eigenvector of $P$ corresponding to the eigenvalue $1$.
Hence there is a single transformation which achieves upper block triangularisation of all row stochastic matrices.
Let $\{ v_2, \ldots, v_n \}$ be a basis for the $n-1$ dimensional subspace
$e^\perp := \{ x \in \R^n \,:\, e^Tx= 0 \}$. 
Then $\{ e, v_2, \ldots, v_n
\}$  is a basis of $\R^n$.
Consider now the transformation matrix $T :=
\begin{bmatrix}
e & v_2 & \ldots & v_n     
\end{bmatrix}$ 
which represents a change of basis from the standard basis to the new basis.
Under this transformation, a row stochastic matrix $P$  is transformed as follows:
\begin{equation}
\label{eq:reparametrization}
T ^{-1}P T =
\begin{bmatrix}
1 & c \\ 0 & Q
\end{bmatrix}
\,.
\end{equation}

%\section{Consensus}
\subsection{Facts about consensus}

Given a sequence of row stochastic matrices $\{ P(k) \}_{k\in\N}$,
consider the time-varying linear system
\begin{equation}
\label{eq:consys}
x(k+1) = P(k) x(k)\,.
\end{equation}
A solution of \eqref{eq:consys} is represented by the left products of
the matrix sequence in the following sense: a sequence $\{ x(k)
\}_{k\in\N}$ is a solution of \eqref{eq:consys} corresponding to the
initial condition $x(0) =x_0$ if and only if
for all $k\in \N$,
\begin{equation}
\label{eq:leftprods}
x(k) =\Phi(k)x_0
\end{equation}
where
\begin{equation}
\Phi(k) := P(k-1) \cdots P(0)
\label{eq:leftprods2}
\qquad \text{for all} \quad k \in \N \,.
\end{equation}
The sequence $\{ P(k) \}_{k\in\N}$ is called {\sf weakly ergodic} if the
difference between each pair of rows converges to zero, i.e. if for all
$i,j$ we have
\begin{equation}
\label{eq:weakergoddef}
\lim_{k\to \infty} \left( e_j^T - e_i^T \right) \Phi(k) = 0 \,.
\end{equation}
This is equivalent to system 
\eqref{eq:consys} being  a 
{\sf consensus system},
that is, 
every solution $\{ x(k)\}_{k\in \N}$ of \eqref{eq:consys} satisfies
\begin{equation}
\label{eq:consensusSysf}
\lim_{k\to \infty} x_j(k)- x_i(k) = 0 
\end{equation}
for all $i,j$.
The sequence $\{ P(k) \}_{k\in\N}$ is {\sf strongly ergodic} if it is weakly ergodic and, in addition,
the limit $\lim_{k\to \infty} \Phi(k)$ exists; we denote this limit by $\Phi_\infty$.
Due to a result of Chatterjee
and Seneta \cite{chatterjee1977towards},  weak and strong ergodicity are equivalent for left
products of row stochastic matrices. 
This is equivalent to 
every solution of \eqref{eq:consys} satisfying
\begin{equation}
\label{eq:consensusLim}
\lim_{k\to \infty} x(k)\in E \,,
\end{equation}
or, equivalently,
\begin{equation}
\label{eq:conscond}
\lim_{k\to \infty} \Phi(k) x_0 \in E 
\end{equation}
for all
$x_0 \in \R^n$ 
where
$
E:= \spann \{ e \}
$ is  the space of consensus vectors. 
We call the sequence $\{ P(k)
\}_{k\in\N}$ 
{\sf strongly ergodic for all initial times}, if all tail sequences $\{ P(k)
\}_{k=k_0}^\infty$ are strongly ergodic for all $k_0 \in \N$. Note that a
sequence can be strongly ergodic and not strongly ergodic for all initial times. For
instance, if one of the matrices in the sequence has rank $1$ and all the
subsequent matrices are the identity matrix.
%
%Note that a sequence can be strongly ergodic and not uniformly strongly ergodic. For
%instance, if one of the matrices in the sequence has rank $1$ and all the
%subsequent matrices are the identity matrix. We will need a particular
%rank one row stochastic matrix, which we denote
%\begin{equation}
%	\label{eq:P1def}
%	\PP := \frac{1}{n}\ e e^\top\,.\newline
%\end{equation}
%
Using the transformation
\eqref{eq:reparametrization} a system  equivalent  to 
\eqref{eq:consys} is given by
\begin{equation}
\label{eq:consystrans}
\begin{array}{rcl}
z(k+1) &=&T^{-1} P(k) T z(k) \\
T^{-1} P(k) T&:=&
\begin{bmatrix}
1 & c(k) \\ 0 & Q(k) 
\end{bmatrix}  \,.
\end{array}
\end{equation}
It is then clear that $ \{ P(k) \}$ is strongly ergodic if and only if
\begin{equation}
\lim_{k \to \infty} Q(k)
\dots Q(0) = 0
\,.
\end{equation}

A useful property in the study of products of row stochastic matrices is the observation that for any row stochastic matrix $P$ we have
\begin{equation}
\label{eq:maxmin}
\textrm{min}(x) \leq \textrm{min}(Px) \leq \textrm{max}(Px) \leq \textrm{max}(x) 
\end{equation}
for all  $x \in \R^n$,
where for any vector $y \in \R^n$,
\[
\textrm{min}(y):= 	\textrm{min}\{ y_1,\dots, y_n \} \,,
\
\textrm{max}(y) :=\textrm{max}\{ y_1,\dots, y_n \} \,.
\]
%Since  the  difference between the maximum and minimum of a vector plays the role of a
%Lyapunov function we introduce the function 
Introducing the function
\begin{equation}
\label{eq:Vdef}
V (x) :=  \textrm{max}(x) - \textrm{min}(x) \,,
\end{equation}
\eqref{eq:maxmin} implies that $V(Px) \leq V(x)$.
%FW: changed to if and only if
Also, the sequence $\{P(k)\}_{k\in \N}$ is strongly ergodic, if and only if
\begin{equation}
\lim_{k \rightarrow \infty} V\big(\Phi(k) x_0\big) = 0
\end{equation}
for all $x_0 \in \R^n$ where $\Phi(k)$ is given by (\ref{eq:leftprods2}).
%In this note the standard norm used is the Euclidean norm
%$\|x\| = \sqrt{x^Tx}$.
% is the maximum norm $\|x \| :=
%\max_{i} | x_i|$ as it has the pleasant property that for all row
%stochastic matrices we have $\|P\| = 1$. 
Note that any vector $x \in \R^n$ can be uniquely decomposed as
\begin{equation}
x = \overline{x} e + x_{\perp}
\end{equation} 
where
\begin{equation}\label{xbardef}
\overline{x} := (1/n)e^Tx 
\end{equation}
is the mean of the components of $x$ and
\begin{equation}
x_\perp := x - \overline{x}e\,,
% \ \in e^{\perp} \,,
\end{equation}
Note that $\overline{x}e \in E$ and $e^{T} x_{\perp} = 0 $;
hence
\begin{equation}
\dist(x, E) = \|x_\perp\|
\end{equation}
where
$
\dist(x,E) := \inf \{ \| x - z \| \,: z \in E \} \,
$
is the distance of a vector $x\in
\R^n $ to the consensus set $E$
and $\|x\| = \sqrt{x^Tx}$.

Note also that
\begin{equation*}
V(x) = V(x_\perp)
\end{equation*}
and
%\begin{equation*}
%\|x_\perp\|_\infty  \le V(x_\perp) \le 2 \|x_\perp\|_\infty
%\end{equation*}
for any vector $z \in \R^n$ and any row-stochastic matrix $P \in \R^{n \times n}$ we have
\vspace{-0.3cm} 
\begin{equation*}
\|z\|_\infty = \max_{1 \leq i \leq n} |z_i| \quad \mbox{and} \quad \|P\|_\infty = \underset{1 \leq i \leq n}{\text{max}} \sum\limits_{j=1}^{n}|P_{ij}| = 1
\end{equation*} 
where $P_{ij}$ is the entry in $i^{\textrm{th}}$ ~row and $j^{\textrm{th}}$ ~column of the matrix $P$.

  As the following results rely on the existence of strongly
  ergodic sequences, it is of course of interest to have criteria for the
  occurrence of such a sequence. This is discussed extensively in the
  literature and here we can only discuss a limited number of references. A number of such criteria can be found in
  \cite{chatterjee1977towards}. Relations of this notion to ergodicity or
 the  Dobrushin coefficient is discussed in \cite{ipsen2011ergodicity} and the
  references given therein. In addition, in the consensus
  literature there are numerous results on the convergence of iterated
  averaging, see e.g. \cite[Theorem 1]{blondel2005convergence}, and
  the survey given in \cite[Section III]{Liter2}.

\section{Consensus under Feedback} \label{Proof}

Consider a sequence of row stochastic matrices $\{
P(k) \}_{k\in\N}$ and a continuous function $G : \R^n \to \R$.  
Then the system,
\begin{equation}
\label{eq:fw-tvsys0}
\begin{array}{rcl}
x(k+1) &= &F(k,x(k))\\
F(k,x) &:=&  P(k) x + G(x) e 
\end{array}
\end{equation}
can be regarded as a consensus system under feedback.
% \begin{equation}
%     \label{eq:fw-tvsys}
%     x(k+1) = P(k) x(k) + f(\overline{x}(k)) e\,.
% \end{equation}
%Here we assume that $\{ P(k) \}_{k\in\N} \subset \R^{n \times n}$ is a
%sequence of matrices in ${\cal P}$ and $f : \R \to \R$ is the feedback.
%
%In a second application we consider continuous functions $f_i : \R \to
%\R$, $i=1,\ldots,n$ and the diagonal map $F : \R^n \to \R^n$ given by
%$x\mapsto \diag \left( f_1(x_1) ,\ldots, f_n(x_n)\right)$. We now consider the
%feedback system
%\begin{equation}
%    \label{eq:fw-tvsys2}
%    x(k+1) = P(k) x(k) + \PP F(x) \,.
%\end{equation}
%
In later statements, further differentiability assumptions will be imposed
on $G$ as required. 
If we apply the similarity transformation defined in
\eqref{eq:reparametrization}, then in the new coordinates,  $y\in \R, z\in \R^{n-1}$,
given by $x = T  [
y  \,\,z^T
 ]^T$,
we obtain
%\begin{equation}
\label{eq:cascade}
\begin{align}
%\begin{array}{rcl}
y(k+1) &= y(k) + G(T   [ y(k) \, z(k)^T ]^T) + c(k) z(k) 				\nonumber\\ 
z(k+1) & = Q(k) z(k) \,.					\label{eq:cascade}
%\end{array}
\end{align}
%\end{equation}
The class of cascaded systems studied in
\cite{loria2002stability,loria2003uniform,nevsic2004uniform} encompasses
this system formulation. We note that in all these references and in
subsequent literature based on these papers \cite{lee2006uniform}, it is assumed that 
the subsystems are globally uniformly asymptotically stable.
We do not require this assumption.
Associated with \eqref{eq:fw-tvsys0} we consider the
one-dimensional system
\begin{equation}
\label{eq:fw-tvsys0_1dim}
\begin{array}{rcl}
y(k+1) &=& h(y(k))\\
h(y) &:=&  y + G(ye) \,,
\end{array}
\end{equation}
which is seen to be the one dimensional subsystem in the cascade
\eqref{eq:cascade} corresponding to $z(k)=0$.
% \begin{equation}
%     \label{eq:fw-tvsys1dim}
%     y(k+1) = y(k) + f(y(k))\,,
% \end{equation}
% resp.
% \begin{equation}
%     \label{eq:fw-tvsys1dim2}
%     y(k+1) = y(k) + \frac{1}{n} e^\top F(y(k)e)
% \end{equation}
This is the aforementioned Lur'e system and,  as we shall see,
the dynamics of the consensus system \eqref{eq:fw-tvsys0} is
strongly related to the dynamics of \eqref{eq:fw-tvsys0_1dim}. Unless
stated otherwise we consider the systems \eqref{eq:fw-tvsys0} and
\eqref{eq:fw-tvsys0_1dim} with initial time $k_0 = 0$. A few comments on
results that hold uniformly with respect to all initial times are made
where appropriate.

\subsection{Local Stability Results}

\begin{lem}
	\label{lem:solspane}
	Let  $\{ P(k) \}_{k\in\N} $ be a sequence of row
	stochastic matrices and $G : \R^n \to \R$.
	If $\{ y(k)
	\}_{k\in\N}$ is a solution of \eqref{eq:fw-tvsys0_1dim} then $\{ y(k) e
	\}_{k\in\N}$ is a solution of \eqref{eq:fw-tvsys0}.
\end{lem}

\begin{proof}
	This follows from $P(k) e = e$.
\end{proof}
The next result tells us that the consensus system under feedback
\eqref{eq:fw-tvsys0} is also a consensus system. We omit the straightforward proof of this observation.
\begin{lem}
	\label{lem:limitconsensus}
	If $\{ P(k) \}_{k\in\N} $ is  a strongly ergodic sequence of row
	stochastic matrices, 
	then for every solution
	$\{ x(k) \}_{k\in\N}$ of \eqref{eq:fw-tvsys0} we have
	\begin{equation}
	\label{eq:limcons}
	\lim_{k\to \infty} \dist \left( x(k), E \right) = 0\,.
	\end{equation}
\end{lem}

% \begin{proof}
% 	The proof of the lemma can be easily shown from \cite{chatterjee1977towards}, \cite[Theorem 1]{blondel2005convergence}, and \cite[Theorem 2]{Liter2}.
% \end{proof}

%\begin{proof}
%	Consider any  solution  $\{ x(k) \}_{k\in\N}$ of \eqref{eq:fw-tvsys0} and let $x_0=x(0)$.
%	Since   
%	$\{ P(k) \}_{k\in\N}$ is strongly ergodic we have
%	%\begin{equation*}
%	$	\lim_{k \to \infty} 
%	V\left( \Phi(k) x_0\right) 
%	= 0, \; 
%	$ where $\Phi(k)$ is given by (\ref{eq:leftprods2}).
%	On the other hand,
%	\begin{align*}
%	V(x(k+1))
%	&=
%	V \left( P(k) x(k) + G(x(k))e \right)\\
%	&=
%	V\left( P(k) x(k)  \right) \,.
%	\end{align*}
%	This shows by induction that for all $k \in \N$ we have
%	%\begin{equation*}
%	$	V(x(k)) =
%	V \left( \Phi(k) x_0\right).$ 
%	%\end{equation*}
%	Hence,
%	%\[
%	$\lim_{k \to \infty} 
%	V\left( x(k)\right) 
%	= 0,$
%	which  is equivalent to
%	$
%	\lim_{k\to \infty} x_j(k)- x_i(k) = 0 
%	$
%	for all $i,j$.
%	This is the same as
%	the desired result \eqref{eq:limcons}.
%\end{proof}

\vspace{0.25cm}

We now consider the local stability of (\ref{eq:fw-tvsys0}) 
%under consensus
and see that it is determined by the stability of the induced system  \eqref{eq:fw-tvsys0_1dim}  on
the consensus space.
As we have no global concerns no Lipschitz property
of $G$ is required.  Initially, it is sufficient that $G$ be continuous.

%In the formulation of the following Lemmas and Theorems it is useful to use the notation $G_e: \R\to \R$, $y \mapsto G(ye)$. 

\begin{thm}{}
	\label{t:localstab}
	Let $\{ P(k) \}_{k\in\N}$ be a strongly ergodic sequence of row
	stochastic matrices and $G:\R^n \rightarrow \R$ be continuous.
	%  and consider systems
	%\eqref{eq:fw-tvsys0} and \eqref{eq:fw-tvsys0_1dim}.  
	Suppose  that
	$y^\ast$ is a locally asymptotically stable fixed point of the one
	dimensional system \eqref{eq:fw-tvsys0_1dim}. Then $y^\ast e$ is a
	locally asymptotically stable fixed point at time $k_0 = 0$ for
	\eqref{eq:fw-tvsys0}.
	
	If the sequence $\{ P(k) \}_{k\in\N}$ is strongly ergodic for all initial times,
	then $y^\ast e$ is asymptotically stable for all initial times $k_0
	\in \N$.
\end{thm}

\begin{proof}
	Suppose that $y^\ast$ is a locally asymptotically stable fixed point for
	system \eqref{eq:fw-tvsys0_1dim}. 
	Let $W$ be a local Lyapunov function which guarantees this stability property.
	That is,  $W(y^*) = 0$ and there is an open neighborhood $U$
	of $y^*$ such  	that $W(y) >0$ and    $W(h(y)) < W(y)$ for all $y\in
	U\setminus \{ y^* \}$. 
	Without loss of generality we may assume $U$ to
	be a forward invariant set of \eqref{eq:fw-tvsys0_1dim}, i.e., if $y \in U$ then $h(y) \in U$.
	%, \forall k \ge k^*$.
	% and introduce the function
	% \begin{equation*}
	%     V(x) := W(\overline x) + V(x) \,.
	% \end{equation*}
	% We claim that $V$ is a local Lyapunov function for the fixed point
	% $y^*e$ of \eqref{eq:fw-tvsys0}. First note that $y^* e$ is clearly a
	% local minimum of $V$. 
	For $\varepsilon >0 $ such that $W^{-1}([0,\varepsilon]) \subset U$ is
	a compact set we may choose $\delta>0$ sufficiently small,
	so that 
	\[
	W(h(y) + d) < \varepsilon
	\qquad  \mbox {for} \qquad
	% $y\leq
	%W^{-1}([0,\varepsilon])$}
	W(y) < \epsilon
	\quad \mbox{and} \quad |d| \leq \delta
	\,.
	\]
	This is possible by continuity of all the
	functions involved and by the decay property of the Lyapunov
	function $W$. 
	
	We note that for any $x \in \R^n$,   
	$Px = P(\overline{x}e + x_\perp) = \overline{x}e + P x_\perp
	$, where $\bar x$ was defined as the mean of the entries of $x\in \R^n $ (recall from \eqref{xbardef}).
	Hence
	\begin{equation}
	\overline{Px} - \overline{x} = \overline{x} + \overline{Px_\perp} - \overline{x} = (1/n)e^TPx_\perp
	\,.
	\end{equation}
	%\[
	%Px -x = P\left(\overline{x}e + x_\perp\right) - \overline{x}e - x_\perp = (P-I)x_\perp
	%\]
	Given a sufficiently small $\varepsilon>0$ and an
	appropriate $\delta$ as above, choosing $\eta>0$ such that
	$V(x) \leq \eta$ and 
	%$\overline{x} \in
	%W^{-1}([0,\varepsilon])$ 
	$W(\bar{x}) \le \epsilon$ implies that for any row stochastic matrix $P$
	\begin{equation*}
	|\overline{Px} - \overline{x}|+ |G(x) - G(\overline{x}e)| 
	< \delta\,.
	\end{equation*}
	This is possible by the estimate for 
	 {\small
        \[
	\|Px - x\|_\infty = \|Px_\perp - x_\perp\|_\infty \leq \|Px_\perp\|_\infty + \|x_\perp\|_\infty \leq 2 \| x_\perp\|_\infty
	\]
	}
	and by uniform continuity of $G$ on a bounded
	neighborhood of $y^* e$. 
	Consider now the neighborhood of $y^*e$ given by
	\begin{equation*}
	N_\varepsilon := \{ x \in \R^n \,:\, \overline{x} \in U,\, W(\overline{x}) < \varepsilon,
	V(x) < \eta \} \,.
	\end{equation*}
	We claim that $N_\varepsilon$ is forward invariant at all times $k\in
	\N$. Indeed, if $x(k) \in N_\varepsilon$, then we obtain 
	\begin{align*}
	\overline x(k+1) 
	&=  \overline{P(k)x(k)} + G(x(k))\\ 
	&= \overline{x}(k) + G(\overline{x}(k)e) + d\\
	&= h(\overline{x}(k)) + d
	\end{align*}
	where $d= \overline{P(k)x(k)} - \overline{x}(k) +G(x(k)) - G(\overline{x}(k)e)$.
	Hence $|d| < \delta$ from which it follows that
	%\begin{align*}
	$	W(\overline x(k+1)) 
	<\varepsilon.$ 
	%\end{align*}
	Referring to the argument in the proof of
	Lemma~\ref{lem:limitconsensus}
	\begin{align*}
	V(x(k+1))= V(P(k)x(k))  \leq V(x(k)) < \eta.
	\end{align*}
	
	As $\varepsilon, \eta$ were arbitrary, this shows stability of $y^* e$. To show local attractivity, let $x_0 \in N_\varepsilon$ for
	$\varepsilon>0$ sufficiently small so that stability holds.
	Note that by Lemma~\ref{lem:limitconsensus} and by
	stability we have
	that $\omega(x_0) \subset U e \subset E$ where $\omega(x_0)$ is the $\omega$-limit set of the solution corresponding to $x_0$ and $Ue: = \left\lbrace ye : y \in U \right\rbrace$ is a subset of $E$.
	Suppose that  $ye
	\in \omega(x_0)$  and 
	$y \neq y^* $ .
	Then as the trajectory starting in $ye$ converges to $y^*
	e$ it follows that $y^* e\in \omega(x_0)$. However, the assumption that 
	$y^* e$ and $ye$ are in the $\omega$-limit set contradicts the stability of
	$y^* e$.
	%The second assertion is immediate.
	Hence $\{x(k)\}_{k \in \N}$ converges to $y^*e$.
\end{proof}

\vspace{0.25cm}

We now extend the previous result to local exponential stability. To this
end we call a sequence of row stochastic matrices $\{ P(k) \}_{k\in\N}$
{\sf exponentially ergodic}  if it is strongly  ergodic and  there exist scalars $M\geq 1$, $0<r<1$  
%and a matrix $\Phi_\infty$ 
such
that for all $k\in \N$
\begin{equation*}
\| \Phi(k) -  \Phi_{\infty}  \| \le  M r^k\,.
\end{equation*}
%FW reformulated the following
The sequence is called  {\sf uniformly exponentially ergodic}, if it is strongly ergodic for all initial times and 
there exists constants $M,r$ so that for all $k_0\in \N$ there exists a matrix $\Phi_{\infty}$ so that 
for all $k>k_0$  we have  $\| \tilde{\Phi}(k,k_0) - \Phi_{\infty} \| \le  M r^{(k-k_0)}$; where $\tilde{\Phi}(k,k_0) := P(k-1) \cdots P(k_0)$.

\begin{thm}{}
	\label{t:localexpstab}
	Let $\{ P(k) \}_{k\in\N}$ be an exponentially ergodic sequence of row
	stochastic matrices and $G:\R^n \rightarrow \R$ be continuously differentiable.
	Suppose  that $y^\ast$ is a locally
	exponentially stable fixed point of the one dimensional system
	\eqref{eq:fw-tvsys0_1dim}.
	Then, $y^\ast e$ is a locally exponentially stable fixed point at time
	$k_0 = 0$ for \eqref{eq:fw-tvsys0}. If the sequence $\{ P(k) \}_{k\in\N}$ is uniformly exponentially
	ergodic, then $y^\ast e$ is locally uniformly exponentially stable.
\end{thm}

\begin{proof}
	Consider the linearisation of the one-dimensional map defining
	\eqref{eq:fw-tvsys0_1dim}.  By the assumption of exponential stability
	it must satisfy
	\begin{equation}
	\label{eq:mod1dim}
	| h'(y^*) | < 1 
	\end{equation}
	where
	$
	h'(y^*) =1 + DG(y^*e)e
	$
	and $DG$ is the derivative of $G$, which we interpret as a row vector.
	We now compute the derivative of $F$ with respect to $x$  at $x=y^* e$ and time
	$k$ to obtain 
	\begin{equation} 
	\label{eq:Jacobian}
	\frac{\partial F}{\partial x}(k, y^* e) = P(k) + e DG(y^*e).
	\end{equation} 
	
	If we now consider the transformation $T$ which results 
	in  \eqref{eq:consystrans}  
	and using $T^{-1} e =
	e_1$  we see that 
	\begin{equation}
	\label{eq:changevar}
	T^{-1} \frac{\partial F}{\partial x}(k, y^* e)T  = \begin{bmatrix}
	1 & c(k) \\ 0 & Q(k) 
	\end{bmatrix} + e_1 DG(y^*e) T\,.
	\end{equation} 
	Two things are noticeable when considering this equation. First the resulting transformed matrix
	is of the form
	\begin{equation}
	\label{eq:changevar2}
	\begin{bmatrix}
	\lambda & \tilde{c}(k) \\ 0 & Q(k) 
	\end{bmatrix}\,,
	\end{equation}
	where only the first row is affected by $G$ and $\lambda$ is independent of
	$k$. 
	Secondly,
	\begin{equation}
	\label{eq:lambda}
	\lambda = 1 + DG(y^*e)e = h'(y^*) \,.
	\end{equation}
	Hence  $|\lambda| < 1$.
	By assumption $\|Q(k) Q(k-1) \dots Q(0)\| \leq M r^{k}$ for
	suitable constants $M\geq 1 $ and $r\in (0,1)$.
	It now follows that the
	linearised system of \eqref{eq:fw-tvsys0} at the fixed point $y^*e$ is
	exponentially stable. It follows by standard linearisation theory
	that the nonlinear system is locally exponentially stable at $y^*e$.
	If the sequence $Q(k) Q(k-1) \dots Q(0)$ converges to
	zero uniformly exponentially, this shows local uniform exponential
	stability of $y^* e$ for the nonlinear system.
\end{proof}

\subsection{Global Stability Results}

To obtain global stability results we first need the following boundedness result.

\begin{lem}
	\label{lem:boundedsols}
	Let $\{ P(k) \}_{k\in\N}$ be a  strongly ergodic sequence of row
	stochastic matrices  
	%Consider systems
	%\eqref{eq:fw-tvsys0} and \eqref{eq:fw-tvsys0_1dim} and ssume that
	and suppose that $G: \R^n\rightarrow \R$ is continuous and satisfies the following conditions.
	\begin{enum} 
		\item There exists an $\varepsilon>0$
		such that $G$ satisfies a Lipschitz condition with constant $L>0$ on
		the set
		\begin{equation*}
		B_\varepsilon(E) := \{ x \in \R^n \,:\, \dist(x,E) \leq \varepsilon \} \,.
		\end{equation*} 
		\item There exist constants $\beta,\gamma > 0$ such that 
		\begin{equation*}
		|h(y)| \leq  |y| - \gamma   \qquad \mbox{when} \quad |y| \geq \beta 
		\end{equation*}
		where $h(y) = y + G(ye)$.
	\end{enum}
	%If \eqref{eq:fw-tvsys0_1dim} has a bounded absorbing set $A$,
	Then every trajectory of \eqref{eq:fw-tvsys0} is bounded.
\end{lem}

\begin{proof}
	%
	%	If one of the matrices $P(k)$ has rank one, say at time $k_0$, then
	%	$x(k) \in \spann \{ e \}$ for all $k\geq k_0$. Then the assertion is
	%	immediate from Lemma~\ref{lem:solspane} as by assumption all solutions
	%	of \eqref{eq:fw-tvsys0_1dim} are bounded. So we may assume without loss of generality that $\rank \Phi(k) > 1$ for all $k$, so that the assumption implies that also all the tail sequences $\{ P(k) \}_{k=k_0}^\infty$ are strongly ergodic for all $k_0$. 
	Consider any solution $ \{ x(k) \}_{k \in \N}$ of \eqref{eq:fw-tvsys0} with $x(0) =
	x_0$. 
	By Lemma~\ref{lem:limitconsensus} there exists a $k_0\in \N$
	such that $x(k) \in B_\varepsilon(E)$ for all $k\geq k_0$. 
	%We will
	%therefore assume from now on without loss of generality that $x(k) \in
	%B_\varepsilon(E)$ for all $k$, so that $G$ has a Lipschitz constant
	%$L$.
	%
	We can express $x(k)$ as	
	%\begin{equation*}
	$x(k) = \overline{x}(k) e + x_{\perp}(k)
	$ %\end{equation*} 
	where 
	$
	\overline{x}(k) = (1/n)e^Tx(k) ~ \mbox{and} ~  x_{\perp}(k) := x(k) -\bar{x}(k)e.
	$
	%Note that $e^{T} x_{\perp}(k) = 0 $. 
	It follows from (\ref{eq:limcons}) that 
	$
	\lim_{k \to \infty}  \| x_{\perp}(k) \| = 0.$
	Hence boundedness of the sequence $\{\bar{x}(k)\}_{k \in \N}$ implies boundedness of $\{x(k)\}_{k \in \N}$.
	Considering  the evolution of $\overline{x}(k)$ we obtain that, for $k\ge k_0$,
	%
	%{\small
		\begin{eqnarray*}
			|\overline{x}(k+1)| &=&
			|\overline{P(k)x(k)}+ G\big(\overline{x}(k)e + x_{\perp}(k) \big) | \\
			&\leq&  |\overline{x}(k) + G(\overline{x}(k)e)| 
			 + |\overline{P(k)x(k)}- \overline{x}(k) | \\
			&+ &
			|G\big(\overline{x}(k)e + x_{\perp}(k)\big) - G(\overline{x}(k)e)|      \\
			&\leq& |h(\overline{x}(k))|  + |(1/n)e^{T}P(k)x_{\perp}(k)|   + L \| x_{\perp}(k) \|\\
			&\le& |h(\overline{x}(k))| + \tilde{L}\| x_{\perp}(k) \|
			\label{eq:upperbound}
		\end{eqnarray*}
	%}
	where
	$
	l := \sup_{k \in \N} (1/n)\|e^TP(k)\|$ and $
	\tilde{L} := l+L$.
	Hence
	\[
	|\overline{x}(k+1)| \le |h(\overline{x}(k))| + \tilde{L}\| x_{\perp}(k) \|
	\,.
	\]	
	It now follows from hypothesis (ii) that whenever $|\bar{x}(k)| \ge \beta$, we must have
	\[
	|\bar{x}(k+1)| \le |\bar{x}(k)| - \gamma  + \tilde{L}\| x_{\perp}(k) \|.
	\]
	%	
	%	&\leq & |\overline{x}(k)| - \gamma + (L+l) \| x_{\perp}(k) \| 
	%
	Since $\lim_{k \to \infty} \| x_{\perp}(k) \| = 0$, 
	there exists a $k_{*} \geq k_0$ such that $\tilde{L} \| x_{\perp}(k) \| \le \gamma$ for all $k > k_{*}$. 
	Thus,
	\[
	|\bar{x}(k+1)| \le |\bar{x}(k)|
	\qquad \mbox{when} \quad k \ge k_* \mbox{ and } |\bar{x}(k)| \ge \beta .
	\]
	This implies boundedness of $\{\bar{x}(k)\}_{k\in \N}$ and  completes the proof.
\end{proof}

\begin{rem} 
As an example of a general class of functions which satisfy  hypothesis (ii) of Lemma \ref{lem:boundedsols}, consider any strict contraction mapping  $h$ on $\R$, i.e., for a suitable constant $c \in (0,1)$,
	\begin{equation*}
	| h(x) - h(y)| \le c | x - y | \,, \quad \forall x,y \in \R .
	\end{equation*}
\end{rem}
By the Banach contraction theorem, there is a unique fixed point $y^*$ such that $h(y^{*}) = y^{*}$. 
Hence,
\begin{align*}
|h(y)| & \leq |h(y) - y^{*} | +  |y^{*}|
  \leq  c |y - y^{*} | +  |y^{*}|\\
&\leq c |y| + (1+c) |y^{*}|
 =|y| -(1-c)|y| +   (1+c) |y^{*}|.
\label{hieq}
\end{align*}
and hypothesis (ii) is assured with  $\beta = \frac{1+c}{1-c}|y^*|$.

Finally, we state a  result on  global asymptotic or exponential
stability. In spirit, the following two results are closely related to
\cite[Theorem 1]{lu2007synchronization}. Note that we obtain a global
result and are only concerned with fixed points, not general
attractors. Also no assumption on the existence or invertibility of the
Jacobian is required. In this sense the result is more general than those
in \cite{lu2007synchronization}. Also strong ergodicity does not imply
uniform asymptotic stability of the $z$-subsystem in the cascade
\eqref{eq:cascade}, therefore the results in
\cite{loria2002stability,loria2003uniform,nevsic2004uniform,lee2006uniform}
are not applicable to the systems considered in  the following theorem.

\begin{thm}{}
	\label{t:globalstab}
	Let $\{ P(k) \}_{k\in\N} \subset \R^{n \times n}$ be a strongly
	ergodic sequence of row stochastic matrices and and suppose that
	$G$ satisfies all conditions of Lemma \ref{lem:boundedsols}.
	If 
	$y^\ast$ is a globally asymptotically stable fixed point of
	\eqref{eq:fw-tvsys0_1dim}
	then, $y^* e$ is a globally asymptotically stable fixed point  for system
	\eqref{eq:fw-tvsys0}.
\end{thm}

%\begin{corollary}
%	\label{c:global}
%	If in Theorem~\ref{t:globalstab} we replace (i) with
%	...
%\end{corollary}

\begin{proof}
	The assumptions of Theorem~\ref{t:localstab} are met and so it only
	remains to show global attractivity.
	Note that, by Lemma~\ref{lem:boundedsols} all
	solutions of \eqref{eq:fw-tvsys0} are bounded. By
	Lemma~\ref{lem:limitconsensus} the $\omega$-limit sets corresponding
	to all initial conditions lie in $E$. 
	So consider an $\omega$-limit
	set $\omega(x_0)$ and  assume  that $ye \in \omega(x_0)$ but $y \neq y^*$.
	Let $U$ be a neighborhood of $y^* e$ on which local stability
	holds according to Theorem~\ref{t:localstab}. We may assume
	$\dist(ye,U) > 0$. As $ye \in E$ it follows
	from  Lemma~\ref{lem:solspane} that all solutions $x(\cdot; k_0,ye)$ with the
	initial condition $x(k_0) = ye$ satisfy
	%	\begin{equation*}
	$		\lim_{k\to \infty} x(k; k_0,ye) = y^*e\,.
	$ %	\end{equation*}
	Note   that on $E$ the system is time-invariant, so that
	there exists a time $K$, such that for all $k_0$ we have
	%	\begin{equation*}
	$		x(k_0+K; k_0,ye) \in U\,.
	$ %	\end{equation*}
	By assumption (i) the maps $x \mapsto P(k)x +  G(x)e$ are
	equicontinuous on $B_\varepsilon(E)$ (i.e., each map, with respect to $k$, is uniformly continuous). Choose $\eta > 0$ such that 
	\begin{equation*}
	B_{\eta,\infty}(E) := \{ x \in \R^n \; ; \;  \dist_\infty(x,E)
	= \min_{r\in \R} \|x - re \|_\infty <\eta \}  
	\end{equation*}
	is contained in $B_\varepsilon(E)$.
	The set $B_{\eta,\infty}(E)$ is forward invariant under all $F(k,\cdot)$,
	because if $\dist_\infty(x,E) = \|x - r_x e \|_\infty <\eta$,
	then as $\|P\|_\infty = 1$ for all row stochastic matrices
	\begin{multline*}
	%\dist_\infty(F(k,x),E) = 
	\dist_\infty(P(k)x + G(x)e, E) 
	%= \dist_\infty(P(k)x, E) \leq \|P(k)x - r_x e \|_\infty =
	\leq \|P(k)(x - r_x e) \|_\infty 
	%\leq  \|x - re\|_\infty 
	<\eta \,.
	\end{multline*}
	% \textcolor{red}{and by \eqref{eq:maxmin} this set is
	% forward invariant under all of these maps.
	% Why is this true? }
	Thus there exists a
	sufficiently small neighborhood $U_2$ of $ye$ such that for all
	$k_0\in \N$ the solution corresponding to the initial condition
	$x(k_0) \in U_2$ satisfies $x(k_0+K;k_0,x(k_0)) \in U$. But then by
	local stability, it follows that $x(k; k_0, x(k_0)) \in U$ for all
	$k\geq k_0+K$. 
	We thus arrive at a contradiction, if $ye \in \omega(x_0)$, then there
	exists a sequence $k_\ell \to \infty$ so that $\lim x(k_\ell; 0, x_0)
	= y e$. But then $x(k_\ell; 0, x_0) \in U_2$ for a sufficiently large
	$\ell$ and hence $x(k;0,x_0) \in U$ for all $k\geq k_\ell+K$. Hence no
	subsequence of $\{x(k)\}$ converges to $ye$.
	This contradiction completes the proof.
\end{proof}

The previous result can be sharpened, if we assume exponential stability
of the fixed point of \eqref{eq:fw-tvsys0_1dim}. We omit the proof, which
uses the same arguments as the proof of Theorem~\ref{t:globalstab} and
which could also be obtained easily by appealing to the methods used in
\cite{loria2002stability}.

\begin{thm}{}
	\label{t:globalstabexp}
	Let $\{ P(k) \}_{k\in\N} \subset \R^{n \times n}$ be a uniformly exponentially ergodic sequence of row
	stochastic matrices and suppose that $G: \R^n\rightarrow \R$ is differentiable and satisfies  conditions (i) and (ii) of Lemma \ref{lem:boundedsols}.
	Then $y^* e$ is globally uniformly exponentially  stable for system
	\eqref{eq:fw-tvsys0}.
\end{thm}

% \begin{proof}
%     The proof follows with the same arguments of the proof of
%     Theorem~\ref{t:globalstab} and is omitted.
% \end{proof}

\subsection{Switched Systems}

Given a compact set of row stochastic matrices ${\mathcal{P}} \subset \R^{n
	\times n}$, we may consider the switched system
\begin{equation}
\label{eq:switchedsys}
x(k+1) = P(k) x(k) + G(x(k)) e\,,
\end{equation}
where $P(k) \in {\mathcal{P}}$.
The results obtained so far have some immediate consequences for consensus
under feedback with arbitrary switching. It is well-known that all
sequences $\{ P(k) \}_{k\in\N} \in {\mathcal{P}}^\N$ are strongly ergodic if and
only if all sequences in ${\mathcal{P}}^\N$ are uniformly exponentially ergodic
\cite{lu2007synchronization}. In this case we call ${\mathcal{P}}$ uniformly
ergodic. The rate of convergence towards $E$ is in
fact given by the projected joint spectral radius
\cite{lu2007synchronization}.

With this in mind the results obtained so far have immediate consequences
for switched systems of the form \eqref{eq:switchedsys}. We note one of
these consequences.

\begin{cor}
	\label{c:switched}
	Let ${\mathcal{P}}$ be a compact set of row stochastic matrices that is
	uniformly ergodic
	and suppose that $G: \R^n\rightarrow \R$ is differentiable and satisfies  conditions (i) and (ii) of Lemma \ref{lem:boundedsols}.
	Then $y^* e$ is  globally uniformly exponentially stable for the switched
	system \eqref{eq:switchedsys} under arbitrary switching. 
\end{cor}

% remark 10 removed. 

%\begin{rem}
%	We have already remarked that this paper is closely related to \cite{martin2} and \cite{lu2007synchronization}. The first of these papers 
%	introduces the present algorithm and shows it performance in simulation studies. However, 
%	the arguments given in that paper do not prove convergence of (4).  
%	The reference \cite{lu2007synchronization} considers linearised dynamics only, whereas the present paper gives a nonlinear and global result for consensus type 	systems with a common input.
%\end{rem}

\section{Optimised Consensus for a Speed Advisory System}

In this section, we describe an application to design a speed advisory system (SAS) for a fleet of vehicles. The objective of this system is  to reduce CO$_{\mbox{\footnotesize{2}}}$ emissions of vehicles running on the highway. Full details of this application are  given in the published paper \cite{ITS}. 
Roughly speaking, we use the idea of optimised consensus, as described above, to calculate a set of virtual speeds, 
which the driver of each  vehicle can use to guide an optimal travelling velocity\protect\footnote{Note that  each driver is still    controlling their vehicle and the speeds are not adjusting automatically. Thus, we do not consider any string stability issues in this simple application.}.

\begin{figure}[htbp]   
	\centering
	\includegraphics[width=0.5\textwidth,height=5cm]{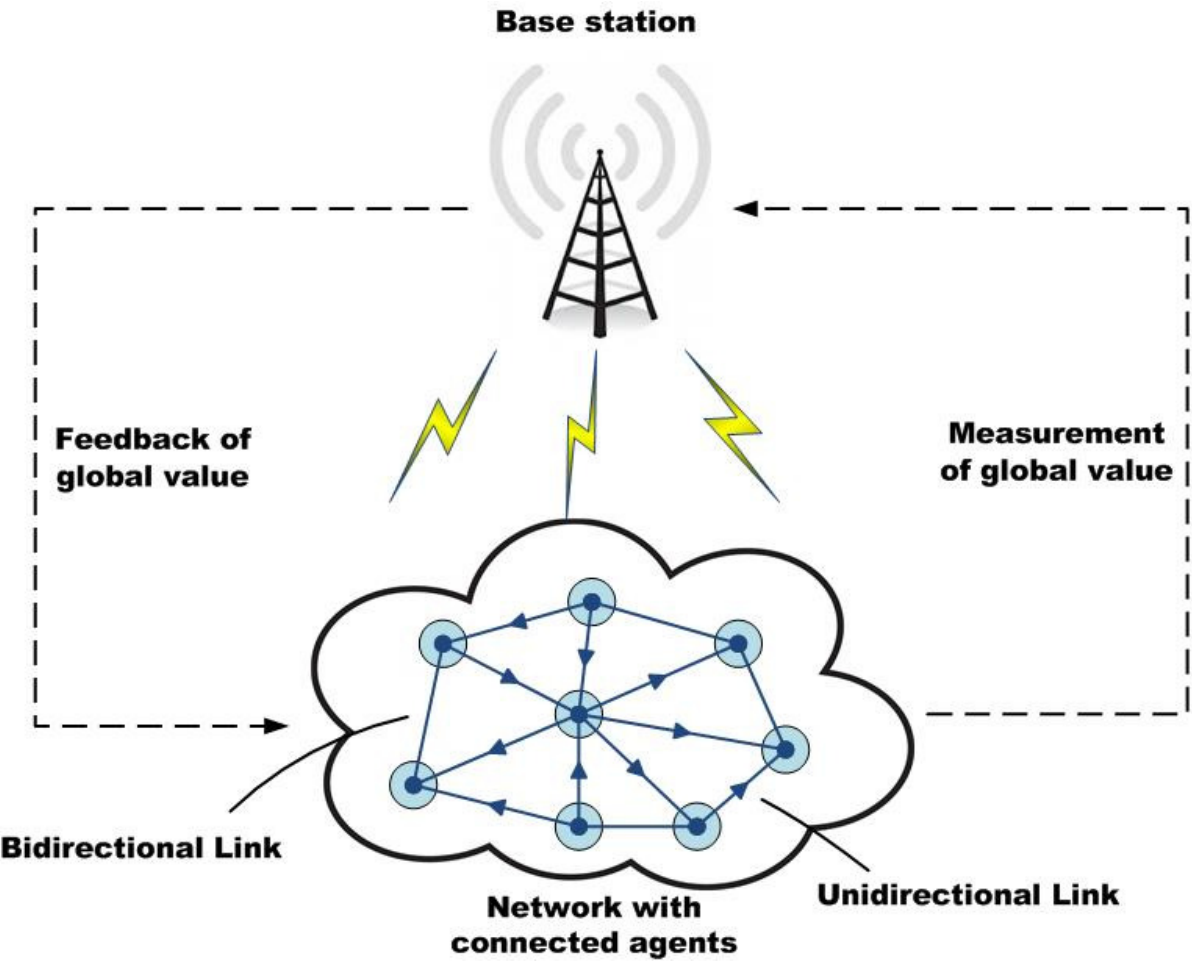}
	\caption{Schematic diagram of the optimisation framework}
	\label{schematic}
\end{figure}

Let us consider a scenario in which a number of vehicles are driving along a given stretch of highway on different lanes in the same direction. Let $N$ denote the total number of vehicles on a particular section of the highway where the SAS broadcast signal can be received. Each vehicle equipped with a specific communication device, which is able to receive and transmit messages to either vehicles or the road infrastructure nearby, is regarded as a mobility agent. Here we define the set $\textrm{\underline{N}} := \left\lbrace 1,2,...,N \right\rbrace$ for indexing the agents. Let $x_i(k)$ denote the recommended speed of the $i^{\textrm{th}}$ ~agent at time slot $k$. The corresponding recommended speed vector for all vehicles at time $k$ is given by $x(k) := \left[x_1(k),x_2(k),...,x_N(k)\right]^{\mathrm{T}}$. In addition, each agent is associated with a CO$_{\mbox{\footnotesize{2}}}$ emission cost function $f_{i}: \mathbb{R} \mapsto \mathbb{R}$, which we assume to be convex, continuous and second order differentiable. We also assume that each agent can adjust $x_i(k)$ based on the knowledge of $f_{i}(x_i(k))$. The first derivative of the $i^{\textrm{th}}$ ~cost function $f_{i}$ is denoted as $f'_{i}: \mathbb{R} \mapsto \mathbb{R}$. In our study, we shall adopt the average-speed model proposed in \cite{emfactor} to model each CO$_{\mbox{\footnotesize{2}}}$ emission cost function $f_i$ as a function of the average speed $x_i$ by
\begin{equation}
	\begin{gathered}
		f_{i}=
		\text{k}\left(\frac{\text{a} + \text{b}x_{i}%\left(k\right)
			+ \text{c}x_{i}^{2}%\left(k\right)
			+ \text{d}x_{i}^{3}%\left(k\right)
			+ \text{e}x_{i}^{4}%\left(k\right)
			+ \text{f}x_{i}^{5}%\left(k\right)
			+ \text{g}x_{i}^{6}%\left(k\right)
		}
		{x_{i}%\left(k\right)
		}\right),
	\end{gathered}
\end{equation}
where $\text{a},\text{b},\text{c},\text{d},\text{e},\text{f},\text{g},\text{k} \in \mathbb{R}$ are used to specify different levels of emissions by different classes of vehicles - see \cite{ITS} for details. Setting of the SAS is depicted in Fig. \ref{schematic}.

The specific mathematical problem we wish to solve is to find an optimal consensus point satisfying $x^*=y^*e$ such that the following optimisation problem is solved:

%\begin{equation} \label{optProblem}
%\begin{gathered}
%\underset{x \in \mathbb{R}^{\textrm{N}}}{\text{minimise}} \quad \sum\limits_{i=1}^{N} f_{i}(x_i)
%\quad \text{subject to: }
%x_i = x_j,  \quad i, j \in \textrm{\underline{N}} \,.
%\\
%{\text{subject to: }} 
%\begin{array}{l}
%x_i = x_j, ~ \forall i, j \in \textrm{\underline{N}} \,.
%\end{array} 
%\end{gathered} 
%\end{equation}

\begin{equation} \label{optProblem}
\underset{x \in \mathbb{R}^{\textrm{N}}}{\text{minimise}} \quad \sum\limits_{i=1}^{N} f_{i}(x_i)
\quad \text{subject to }
x_i = x_j, \,\,\, i, j \in \textrm{\underline{N}} \,.
\end{equation}

After finding this common suggested speed, drivers are encouraged to drive at this recommended speed to minimise group emissions. Note in this study we use COPERT emission functions \cite{emfactor}, but it is also possible to measure these average-speed functions in the car so as to incorporate individual driver behaviour. 

In what follows, we wish to use an iterative feedback scheme of the form
\eqref{eq:fw-tvsys0} to solve the optimisation problem  \eqref{optProblem}.
We will require that this problem has  a unique solution and derive
the specific form for $G$ in \eqref{eq:fw-tvsys0} from first order
optimality conditions.
To this end, it follows from elementary
optimisation theory that when the $f_i$'s are strictly convex, the
optimisation problem will be solved if and only if there exists a unique
$y^{*}\in \R$ such that $
\sum_{i = 1}^{N} f'_{i}(y^{*}) = 0 .
$
%where $f_i'$ denotes the first derivative of the utility function $f_i$.
With
this in mind we apply a feedback signal 
$
G(x) = -\mu \sum_{i = 1}^{N}
f'_{i}(x_{i})$
where $\mu \in \R$ is a parameter to be determined. This gives rise to the following dynamical system 
\begin{equation} \label{optNew}
x(k+1) = P(k)x(k) -\mu \sum_{i = 1}^{N} f'_{i}(x_{i}(k)) e
\end{equation}
where for each $k$ we define $P(k)$ as  

\begin{equation} \label{Pk}
P_{{i} {j}}\left(k\right)=\left\{ \begin{array}{cl}
\displaystyle{1-\sum_{{j}\in N_{k}^{{i}}}\eta_{{j}} } & \mbox{if }{j}={i}\\
\eta_{{j}} & \mbox{if }{j}\in N_{k}^{{i}}\\
0 & \mbox{otherwise}
\end{array}\right.
\end{equation}
where 
%${i}$, ${j}$ are the indices of the entries of the matrix $P\left(k\right)$, 
$\eta_{{j}} \in \mathbb{R}$ is a weighting factor, and $N_k^{{i}}$ represents the set of neighbour agents communicating to the $i^{\textrm{th}}$ vehicle. 

As we assume that the $f_i$ are strictly convex, their derivatives are strictly increasing. We
assume that each $f'_{i}$ has a strictly positive and bounded growth,
i.e., there exist constants  $d_{\min}^{(i)}$ and $d_{\max}^{(i)}$; such
that  for any $a \neq b$
\begin{equation} \label{bdd}
0 < d_{\min}^{(i)} \leq \frac{f'_{i}(a) - f'_{i}(b)}{a - b} \leq d_{\max}^{(i)}\qquad  \forall i \in \textrm{\underline{N}}. 
\end{equation}
% \noindent Define further 
% \begin{equation}
% \begin{gathered}
% g_{\min} = \sum\limits_{i=1}^n \frac{d_{\min}^{(i)}}{d_{\max}^{(i)}} , ~g_{\max} = \sum\limits_{i=1}^n \frac{d_{\max}^{(i)}}{d_{\min}^{(i)}}.
% \end{gathered}
% \end{equation}
We claim that provided $\mu$ is chosen according to 

$0 < \mu < 2 \left( \sum\limits_{i=1}^N d_{\max}^{(i)} \right)^{-1}$
then (\ref{optNew}) is uniformly globally asymptotically stable
at the unique optimal point $x^* e$ of  the optimisation problem
(\ref{optProblem}). First, we consider the scalar system associated with (\ref{optNew}) which is given by
\begin{equation} \label{yNew}
y(k+1) = y(k) -\mu \sum_{i = 1}^{N} f'_{i}(y(k)).
\end{equation}
Note first that the fixed point condition for \eqref{yNew} is $\sum_{i = 1}^{N} f'_{i}(y^*) = 0$. So that a fixed point $y^*$ of \eqref{yNew}, gives rise, 
by Lemma~\ref{lem:solspane} to a fixed point of \eqref{optNew}, which
satisfies the necessary and sufficient conditions for optimality. Now, we wish to use Theorem~\ref{t:globalstab} to show global asymptotic
stability.  To this end, we need to verify that system
(\ref{optNew}) satisfies all the conditions required in
Theorem~\ref{t:globalstab}. The condition on $\mu$ ensures that the right hand side of \eqref{yNew} is in fact a strict contraction on $\R$. It follows from our comments after Lemma \ref{lem:boundedsols} that the assumption  (ii) of Lemma \ref{lem:boundedsols} is satisfied. To show the Lipschitz condition
(i) note that by \eqref{bdd} each $f'_i$ is globally Lipschitz. As the
coordinate functions are globally Lipschitz and sums of globally Lipschitz
functions retain that property we obtain condition (ii). 

\begin{figure}[htbp]   
	\hspace{-0.4cm}
	\includegraphics[width=0.50\textwidth,height=8cm]{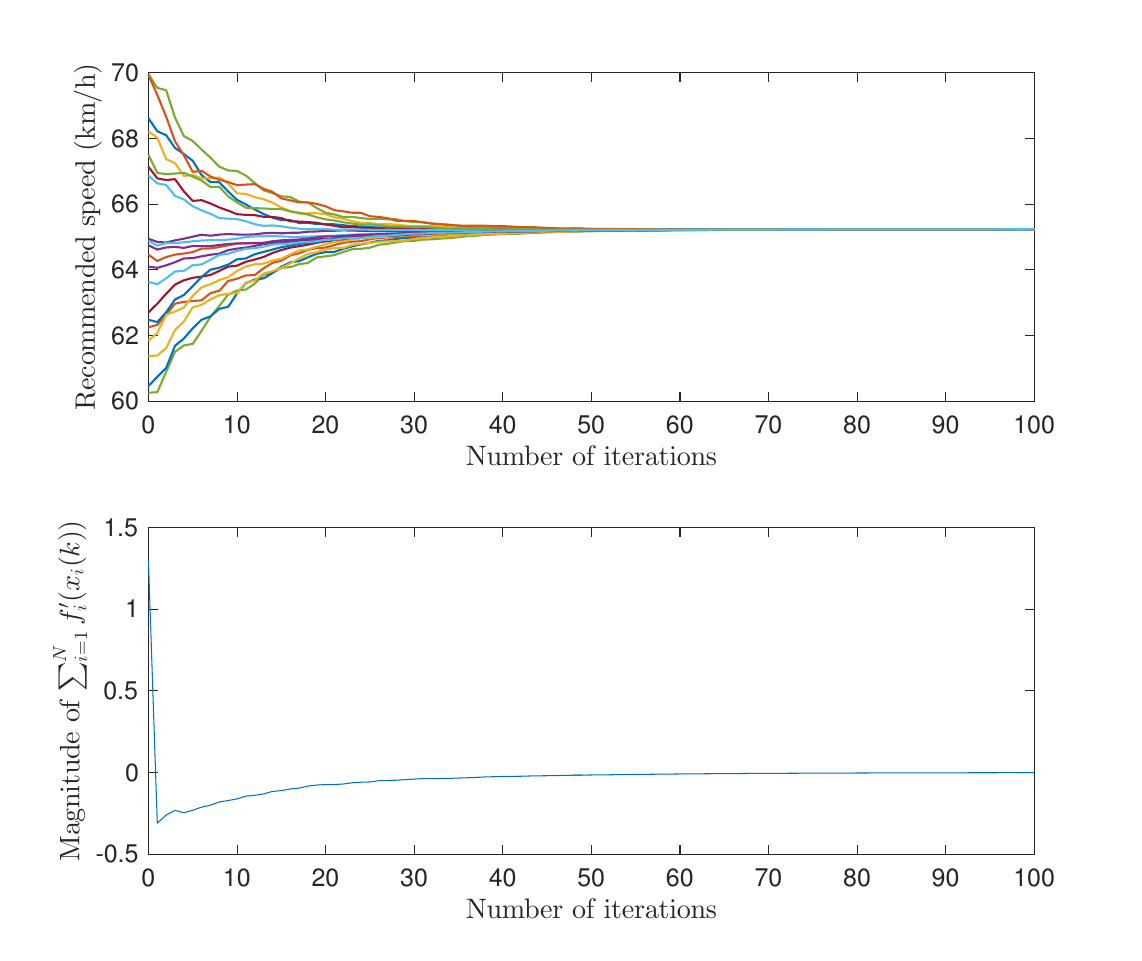}
	\caption{Dynamics of the state variables $x(k)$ with $\mu=0.1$}
	\label{simResults}
\end{figure}

To illustrate this application we consider 20 vehicles travelling along a section of road with random initial suggested speeds uniformly distributed between 60km/h and 70km/h. We assumed that there were 10 vehicles for each emission class, and the parameters set of the cost function for each class was chosen as R007 and R104, respectively, from \cite{ITS}. The simulation results are presented in Fig. \ref{simResults}. Our results show that the recommended speeds of vehicles will asymptotically converge to the optimal one in less than 100 algorithm iterations.  
 
\vspace{1cm}
\begin{rem}
	Our main results are directed at consensus type applications where a basic type of ergodicity is assumed to hold. Clearly, this assumption is not always true. 
	However, we note the following facts which are pertinent for applications in intelligent transportation systems (ITS), each of which make the assumption of strong ergodicity plausible.\newline
	\begin{itemize}
	\item[(i)] We are primarily motivated by ITS applications in which agents (cars) are travelling in close proximity to each other, thereby giving rise to connected communication graphs \cite{ITS}.  In areas of sparsely connected cars, roadside infrastructure can play a role in controlling the connectivity of the graph.\newline
	\item[(ii)] Many  applications of this type also operate a form of topology control to ensure either spatial or temporal connectivity. Details of one such algorithm is given in \cite{martin2,knorn2011results}.\newline
	\item[(iii)] Any real SAS system will almost certainly be augmented by a local vehicle-based vision system. This vision system can provide estimates of neighbouring vehicle speeds, and these can be used as a proxy for suggested speeds. Also, if drivers do not follow the suggested speed, and others do, then vehicles will become closer in space to each other, thereby making the graph more connected, and this will have the effect of making the graph strongly ergodic.\newline 
	\item[(iv)] Finally, the central authority can be used to send global information (other than derivatives), every so often, so as to make strong ergodicity even more likely. 	
	\end{itemize}
\end{rem}

\section{Conclusion}  \label{Conclusion}
In this note we present a rigorous proof of stability and convergence of a recently proposed consensus system with feedback. Examples are given to illustrate the usefulness of the algorithm. For other smart grid applications see \cite{IEVC2014}.

\begin{ack}                               % Place acknowledgements
This work was in part supported by Science Foundation Ireland under grant 11/PI/1177.
\end{ack}

\bibliographystyle{plain}        % Include this if you use bibtex 
\bibliography{References}           % and a bib file to produce the 
                                 % bibliography (preferred). The
                                 % correct style is generated by
                                 % Elsevier at the time of printing.

\vspace{1cm}

\parpic{\includegraphics[width=1in,clip,keepaspectratio]{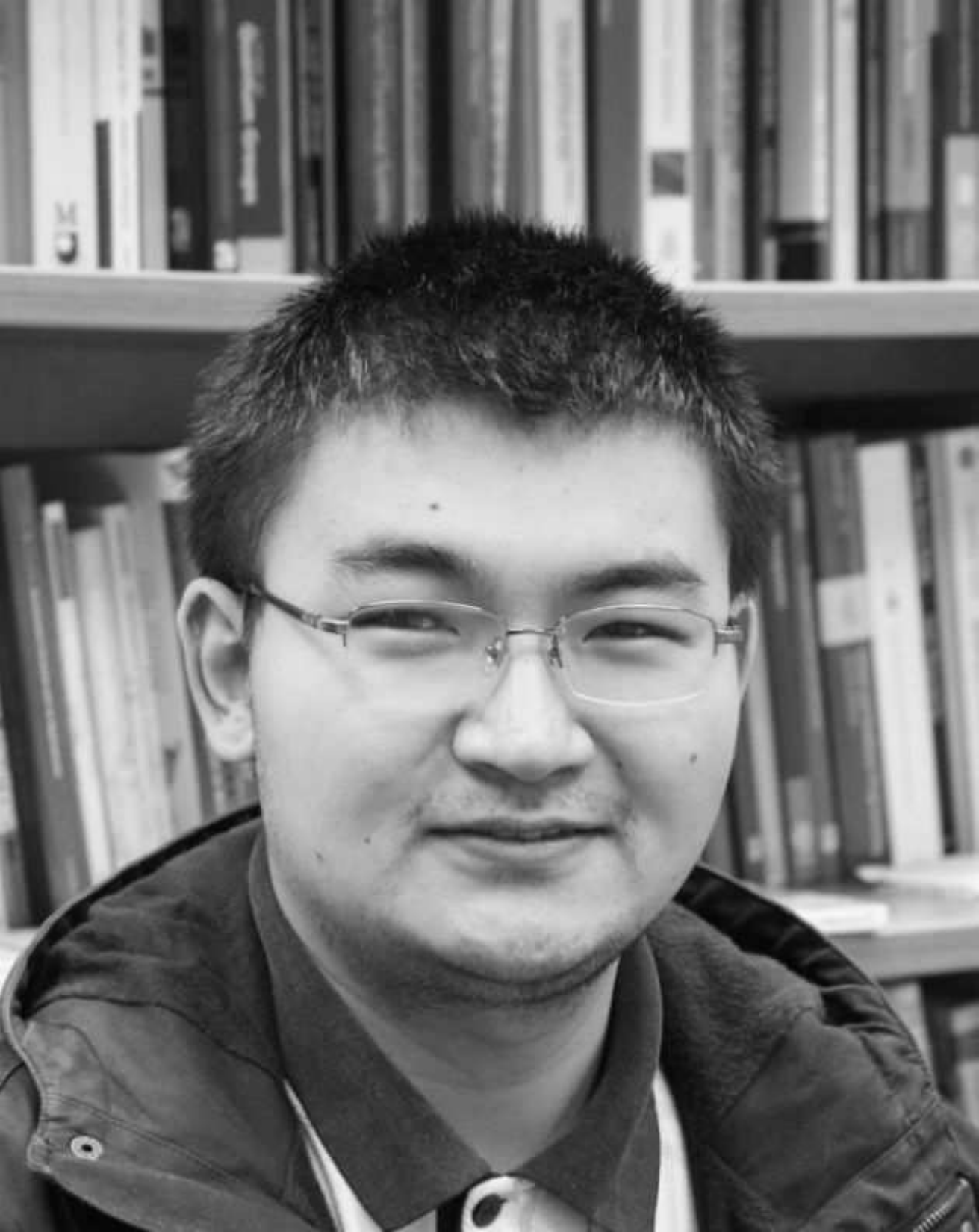}}
\noindent {\bf Mingming Liu} received his B.E. and Ph.D. degrees from National University of Ireland Maynooth in 2011 and 2015 respectively. He is currently a Post-Doctoral Research Fellow with University College Dublin. His research interests are nonlinear system dynamics, distributed control and optimisation techniques with applications to smart grid and transportation systems.

\parpic{\includegraphics[width=1in,clip,keepaspectratio]{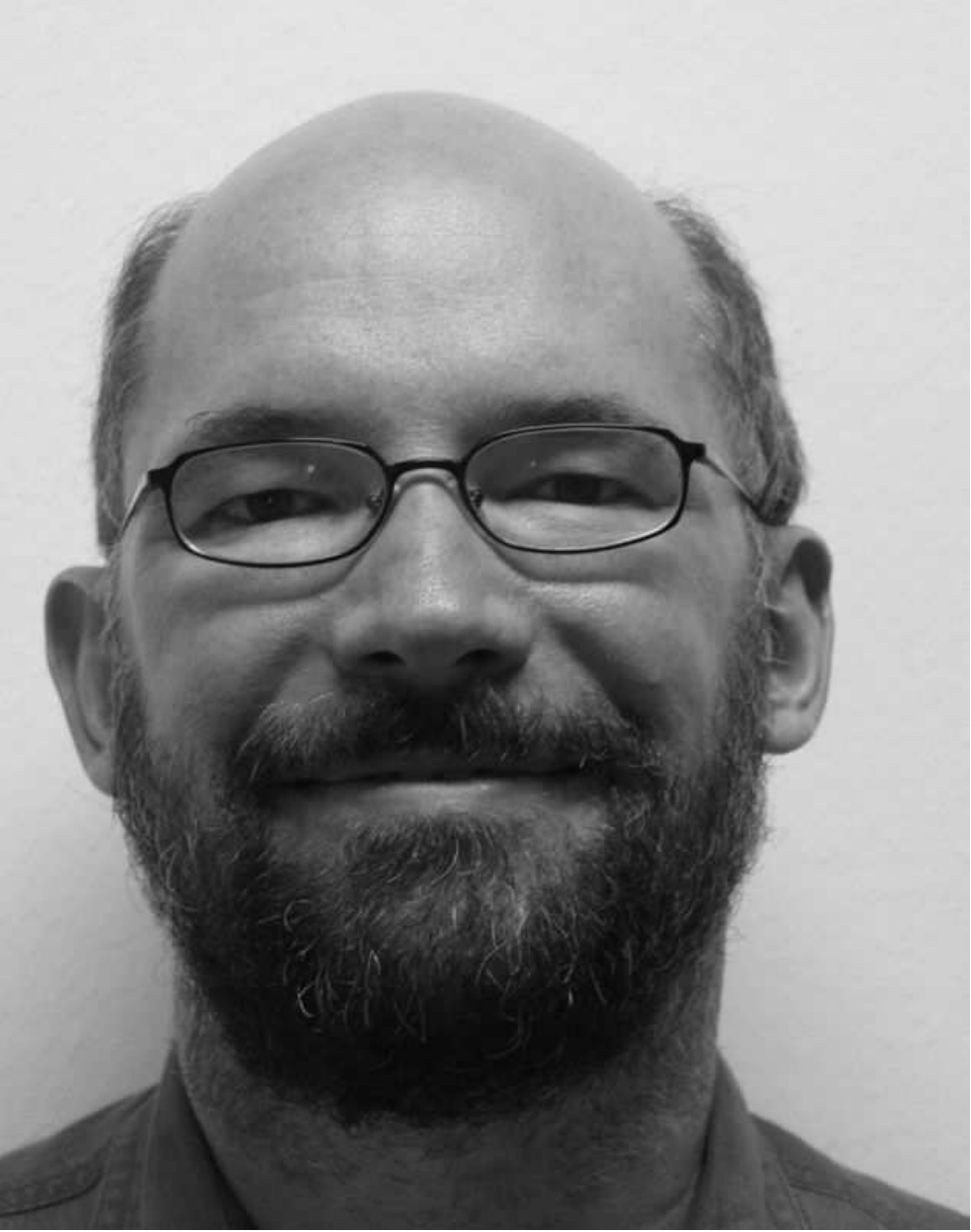}}
\noindent {\bf Fabian Wirth} received his PhD from the Institute of Dynamical Systems at the University of Bremen in 1995. He has since held positions in Bremen, at the Centre Automatique et Syst{\`e}mes of Ecole des Mines, the Hamilton Institute at NUI Maynooth, Ireland, the University of W\"{u}rzburg and IBM Research Ireland. He holds the chair at the University of Passau, Germany. His current interests include stability theory, queueing networks, switched systems and large scale networks with applications to networked systems and in the domain of smart cities.

\parpic{\includegraphics[width=1in,clip,keepaspectratio]{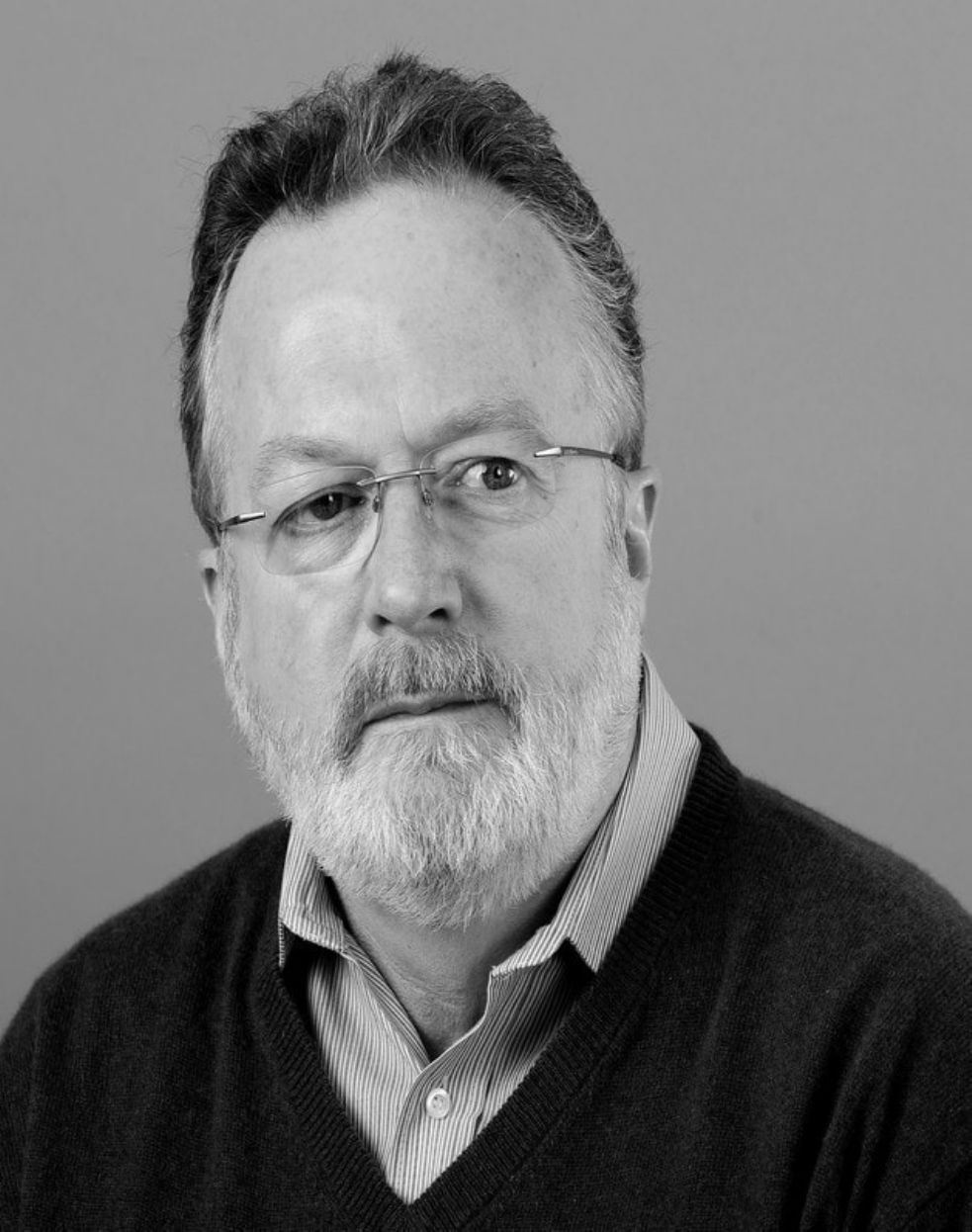}}
\noindent {\bf Martin Corless} is  a Professor in the School of Aeronautics \& Astronautics at Purdue University. He is also Visiting Professor at University College Dublin and an Adjunct Honorary Professor at The National University of Ireland, Maynooth. He received a B.E. from University College Dublin and a Ph.D. from the University of California at Berkeley; both degrees are in mechanical engineering. His research is concerned with obtaining tools which are useful in the robust analysis and control of systems containing significant uncertainty and in applying these results to aerospace and mechanical systems and to sensor and communication networks.

\parpic{\includegraphics[width=1in,clip,keepaspectratio]{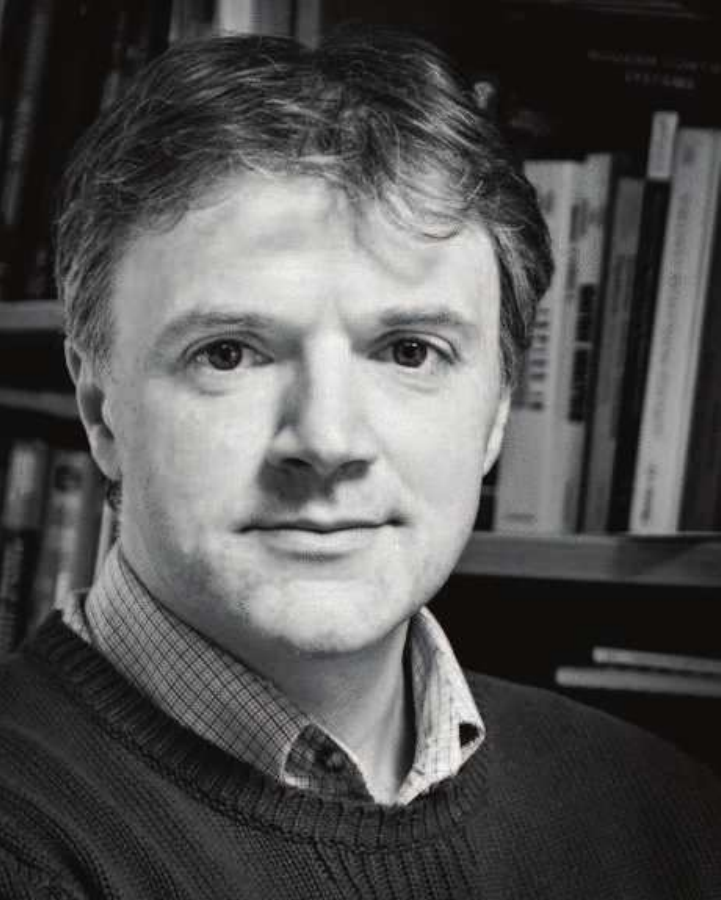}}
\noindent {\bf Robert Shorten} is currently Professor of Control Engineering and Decision Science at University College Dublin. Prof. Shorten's research spans a number of areas. He has been active in computer networking, automotive research, collaborative mobility (including smart transportation and electric vehicles), as well as basic control theory and linear algebra. His main field of theoretical research has been the study of hybrid dynamical systems.

\end{document}